# Photo-induced Magnetic Force Between Nanostructures

Caner Guclu, Venkata Ananth Tamma, Hemantha Kumar Wickramasinghe, Filippo Capolino
Department of Electrical Engineering and Computer Science, University of California, Irvine, CA 92697

*Abstract*— Photo-induced magnetic force between nanostructures, at optical frequencies, is investigated theoretically. Till now optical magnetic effects are not used in scanning probe microscopy because of the vanishing natural magnetism with increasing frequency. On the other hand, artificial magnetism in engineered nanostructures led to the development of measurable optical magnetism. Here, two examples of nanoprobes that are able to generate strong magnetic dipolar fields at optical frequency are investigated: first an ideal magnetically polarizable nanosphere and then a circular cluster of silver nanospheres that has a loop-like collective plasmonic resonance equivalent to a magnetic dipole. Magnetic forces are evaluated based on nanostructure polarizabilities, i.e. induced magnetic dipoles, and magnetic-near field evaluations. As an initial assessment on the possibility of a magnetic nanoprobe to detect magnetic forces, we consider two identical magnetically polarizable nanoprobes and observe magnetic forces in the order of piconewtons thereby bringing it within detection limits of conventional atomic force microscopes at ambient pressure and temperature. The detection of magnetic force is a promising method in studying optical magnetic transitions that can be the basis of innovative spectroscopy applications.

## I. Introduction

Magnetic properties of materials are very weak and almost completely vanish at optical frequencies [1]. Therefore, matter's magnetic properties and magnetic field-matter interactions at optical frequencies are treated as exotic and their investigation is in general absent in science and engineering. Since matter also possesses electronic states that involve circulation of electrons and spin transitions at optical frequencies, it would be desirable, despite their weakness, to be able to directly interact with magnetic dipole transitions.

There has been recent interest in optical-frequency magnetic response of engineered nanostructures such as photonic crystals and optical metamaterials [2]–[4]. Artificial magnetism is achieved at optical frequencies by enhancing and controlling the emission characteristics of the magnetic dipolar term in the multipolar expansion of scattered fields [2], [5]–[13] by nanostructures. Therefore, metamaterial-based magnetic nanostructures open up new possibilities of exploring the optical magnetic field - matter interactions. In our pursuit to develop future tools to investigate optical magnetic interactions, it is of particular interest to explore the concept of photo-induced *magnetic* nanoprobes (i.e., working at optical frequencies) that can be used in scanning probe microscopy (SPM) [14]–[18]. Some recent research focused on the enhancement of magnetic field near nano-apertures to record the spatial distribution of magnetic fields within magnetic nanostructures [19], [20]. In this work, we elaborate on the idea of a magnetic nanoprobe that can be used in SPM to detect photo-induced magnetic forces. The study of optical magnetism through SPM tools based on magnetic nanoprobes can boost the development of spectroscopy applications based on magnetic dipolar transitions.

The capability of atomic force microscopy (AFM) has been extended recently by allowing for measurement of photo-induced forces [21]–[24]. Indeed, it has been experimentally demonstrated that optical forces, measured by an AFM, could be used to image optical electric fields [23] and characterize both linear [21] and non-linear [22] polarizability of materials. While the previous work has been devoted to measuring the optical force between photo-induced electric dipoles [21], [22], [25], [26], in this work, we explore the new concept of nanoprobes capable of supporting photo-induced magnetic polarizability such that the interaction between nanoprobe and matter could be investigated in the future using photo-induced magnetic forces. This concept is fundamentally different from the previously developed magnetic force microscopy [27]–[29] that operates with magnetostatic fields, since here the magnetic force of interest is generated by magnetic field oscillating at optical frequency.

In order to boost the optical magnetic force to measurable levels, a strong magnetic near-field and its spatial derivatives are to be created using a magnetically polarizable probe. As natural magnetism fades at optical frequencies, such a probe design has to rely on artificial magnetism achieved by optical metamaterials. Indeed, optical forces in engineered nanostructures are a subject under of recent attention [30]. Recently, magnetic force on magnetodielectric particles is investigated using Mie polarizabilities [31]–[33].

In particular, in this paper we investigate the electromagnetic interaction between two nanoprobes at nanometer-scale distance from each other illuminated by a LASER beam. We show that a magnetic nanoprobe could be conceived by two exemplary realizations of magnetically polarizable nanostructures: (i) a nanosphere of dense dielectric material supporting a strong magnetic dipolar polarizability; (ii) a circular cluster of plasmonic nanospheres. The force between two identical magnetic nanoprobes driven by a beam is investigated as a measure of possible achievable force levels. In both cases, the force between the two magnetic probes is shown to be in the measurable range even at ambient pressure and temperature. These findings open the way to investigate magnetic dipolar transitions in matter using SPM with specially conceived photo-induced magnetic nanoprobes.

Though not developed in this paper, to further isolate the magnetic force detected by a magnetic nanoprobe from the usually-stronger electric one, structured light illumination could be used, an example of such an illumination scheme is a vector vortex beam with high magnetic-to-electric field contrast along the beam axis as reported in [34], [35].

## II. Force Between Two Magnetically Polarizable Nanoparticles at Optical Frequency

The force exerted by a local magnetic field $\hat{\mathbf{H}}^{\text{loc}}$ on a nanoparticle with magnetic dipole moment $\hat{\mathbf{m}}$ is found by $\hat{\mathbf{F}}(t) = \mu_0 \hat{\mathbf{m}} \cdot \nabla \hat{\mathbf{H}}^{\text{loc}}$ which is a compact representation of $\hat{\mathbf{F}}(t) = \mu_0 \sum_j \hat{m}_j \partial \hat{\mathbf{H}}^{\text{loc}} / \partial j$ with $j = x, y, z$ and $\hat{m}_j$ representing the components of moment $\hat{\mathbf{m}}$ (note that the bold vectors here denote instantaneous quantities and the hat "^" denotes a time domain quantity). In the following we consider the setup in Fig. 1, where the force is only z-directed owing to the symmetry (thus in this section we drop the subscript z in force given by $\hat{\mathbf{F}} = \hat{F} \mathbf{1}_z$) and it is equal to $\hat{F}(t) = \mu_0 \hat{m} \partial \hat{H}_z^{\text{loc}} / \partial z$ with $\hat{\mathbf{m}} = \hat{m} \mathbf{1}_z$ (due to symmetry, the subscript z of magnetic dipole moment z-component is suppressed). A more general case in given in the Appendix. We consider a time-harmonic external field therefore the magnetic dipole and field are expressed as

$$\hat{m} = \text{Re}\{m e^{-i\omega t}\}, \quad \hat{H}_z^{\text{loc}} = \text{Re}\{H_z^{\text{loc}} e^{-i\omega t}\} \quad (1)$$

in terms of complex valued time harmonic vectors (phasors). Hence, the time-average force on a magnetic dipole is found by

$$\langle \hat{F} \rangle = \frac{\mu_0}{2} \text{Re}\left\{ m \left( \frac{\partial H_z^{\text{loc}}}{\partial z} \right)^* \right\}, \quad (2)$$

where the asterisk denotes the complex conjugation.

We consider first a general formulation with two nanoparticles with assigned magnetic polarizability. Then we exemplify it with two possible examples of magnetic nanoprobes either nanospheres with a magnetic polarizability at optical frequencies (for example obtained by a magnetic Mie resonance of a dense dielectric nanosphere), or a circular cluster of plasmonic nanospheres that has been shown to support a loop-like resonance equivalent to a magnetic dipole at optical frequencies [6], [7], [9], [12], [13].

Considering a system of two nanoparticles as in Fig. 1, where each, at position $\mathbf{r}_i$, possesses a magnetic dipole moment $\mathbf{m}_i$, with $i = 1, 2$. Each particle's magnetic moment is related to the local magnetic field $\mathbf{H}^{\text{loc}}(\mathbf{r}_i)$ acting on it by $\mathbf{m}_i = \alpha_i^{mm} \mathbf{H}^{\text{loc}}(\mathbf{r}_i)$, where $\alpha_i^{mm}$ is its magnetic polarizability that is assumed to be isotropic. The local magnetic field acting on a nanoparticle is the sum of the magnetic field generated by the other particle and the external driving field which is assumed slowly varying in space for simplicity, as it would be in several illuminating systems due to the subwavelength distance d and particle radii.

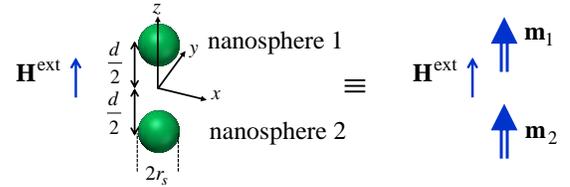

Fig. 1. Two magnetically polarizable nanoparticles excited by an external magnetic field oscillating at optical frequency. A force will be established between the two particles, denoted here as "magnetic force".

The former is evaluated through the dyadic Green's function (GF) $\underline{\mathbf{G}}^{Hm}(\mathbf{r} - \mathbf{r}')$ used for evaluating the magnetic field generated at $\mathbf{r}$ by a magnetic dipole at $\mathbf{r}'$. The system of equations can be constructed as

$$\mathbf{m}_i = \alpha_i^{mm} \mathbf{H}^{\text{loc}}(\mathbf{r}_i) = \alpha_i^{mm} \left[ \mathbf{H}^{\text{ext}} + \underline{\mathbf{G}}^{Hm}(\mathbf{r}_i - \mathbf{r}_j) \cdot \mathbf{m}_j \right] \quad (3)$$

where $i, j = 1, 2$. The external magnetic field is here assumed to be polarized along the z-axis and as a consequence of symmetry the resulting magnetic dipoles will be also purely z-directed. When the two particles are assumed identical, i.e., having the same polarizability, the solution for each magnetic dipole moment reduces to

$$m = \frac{\alpha^{mm} H_z^{\text{ext}}}{1 - \alpha^{mm} G_{zz}^{Hm}(d)} \quad (4)$$

where $G_{zz}^{Hm}$ is the zz entry of the dyadic GF $\underline{\mathbf{G}}^{Hm}$ and is a function of $d = |\mathbf{r}_1 - \mathbf{r}_2|$ only. The evaluation of the force on nanosphere 1 as in (2), is then completed by taking the derivative of the local magnetic field $H_z^{\text{loc}} = H_z^{\text{ext}} + G_{zz}^{Hm}(z + d/2) m$ with respect to z. As mentioned previously, assuming that $H_z^{\text{ext}}$ is slowly varying with z over the system scale, one has $\partial H_z^{\text{loc}} / \partial z = m \partial G_{zz}^{Hm} / \partial z$

and the time-average force takes the form

$$\langle \hat{F} \rangle = \frac{\mu_0}{2}|m|^2 \operatorname{Re}\left\{ \frac{\partial G_{zz}^{Hm}\left(z+\frac{d}{2}\right)}{\partial z} \right\}_{z=\frac{d}{2}}. \quad (5)$$

In the quasi-static regime $d \ll \lambda_0$ (where $\lambda_0$ is the free space wavelength), the Green's function is well approximated as a real function $G_{zz}^{Hm}(z) = 1/(2\pi|z|^3)$. Under this condition we obtain

$$\left.\frac{\partial H_z^{loc}}{\partial z}\right|_{z=\frac{d}{2}} = m \left.\frac{\partial G_{zz}^{Hm}\left(z+\frac{d}{2}\right)}{\partial z}\right|_{z=\frac{d}{2}} = -3m\frac{G_{zz}^{Hm}(d)}{d}. \quad (6)$$

Neglecting the GF dynamical terms, the time-average force on nanosphere 1 is finally represented as

$$\langle \hat{F} \rangle = -\frac{3\mu_0}{2}\left|\frac{\alpha^{mm}H_z^{ext}}{1-\alpha^{mm}G_{zz}^{Hm}(d)}\right|^2 \frac{G_{zz}^{Hm}(d)}{d}. \quad (7)$$

An approximation of the force expression is obtained by neglecting the other dipole's contribution to the local magnetic field where $H_z^{loc} \approx H_z^{ext}$ and hence each magnetic dipole is simply generated by the external magnetic field as $m \approx \alpha^{mm}H_z^{ext}$. In other words the coupling mechanism for the determination of the dipolar strengths is neglected, and this is equivalent to assuming $|\alpha^{mm}G_{zz}^{Hm}(d)| \ll 1$ in the denominator of (4). The approximate expression for the force under this approximation will be noted as $\langle \hat{F}_a \rangle$ and it is given by

$$\langle \hat{F}_a \rangle = -\frac{3\mu_0}{2}|\alpha^{mm}H_z^{ext}|^2 \frac{G_{zz}^{Hm}(d)}{d}. \quad (8)$$

It is clear that the latter scales as $d^{-4}$. In both expressions (7) and (8) it is clear that the force is linearly proportional to the square of the external magnetic field magnitude. Considering an external magnetic field with the amplitude of $2.65\times 10^3$ A/m corresponding to that of an incident beam with an electric field equal to $10^6$ V/m (i.e, with average power density of $1.33\times 10^5$ W/cm$^2$), we report in Fig. 2(a) and (b) the values of the time-average force exerted on nanosphere 1, based on $\langle \hat{F} \rangle$ and $\langle \hat{F}_a \rangle$ in (7) and (8), respectively, versus distance $d$ (horizontal axis) and magnetic polarizability magnitude $|\alpha^{mm}|$ (vertical axis), as a color map. One can observe that for distances $d$ smaller than 250 nm, polarizabilities in the order of $10^{-21}$ m$^3$ are sufficient for realizing a magnetic force in the order of 0.1 pN. The formula in (7) provides a sharp peak due to the feedback (coupling) term $\alpha^{mm}G_{zz}^{Hm}(d)$ in the denominator, whereas the approximated formula (8) provides a monotonical growth of force with decreasing distance. For smaller distances $d$ the coupling term $\alpha^{mm}G_{zz}^{Hm}(d)$ tends to grow as $d^{-3}$ thus $|m|^2$ in (5) decays as $d^6$ which is faster than the growth of $\partial G_{zz}^{Hm}/\partial z$ as $d^{-4}$. As a result, force $\langle \hat{F} \rangle$ decays as $d^2$ for very small $d$, whereas $\langle \hat{F}_a \rangle$ grows with the rate of $\partial G_{zz}^{Hm}/\partial z$, as $d^{-4}$. Therefore the approximate force formula yields miscalculated huge force levels for small distances, and it is vital to include the coupling term.

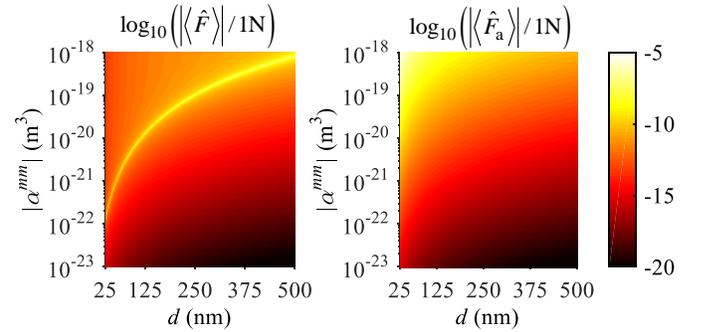

Fig. 2. Plots of magnetic force at optical frequency: (a) $\log_{10}(|\langle \hat{F} \rangle|/1\mathrm{N})$ and (b) $\log_{10}(|\langle \hat{F}_a \rangle|/1\mathrm{N})$, normalized to 1 Newton, versus center-to-center inter-particle distance and the magnetic polarizability of the particles, using quasistatic Green's function.

The above results provide an insight into the required magnetic polarizability to have forces in the order of 0.1 pN; we examine next two magnetic nanoprobes satisfying such requirement. Our first nanoprobe type is a magnetic polarizable nanosphere, that can be obtained for example with nanospheres with large permittivity able to exhibit a magnetic dipole-like Mie resonance even in subwavelength dimensions. For example the magnitude of magnetic Mie polarizability based on formulas 4.56 and 4.57 at page 101 in [36], [37] of a Si nanosphere is shown in Fig. 3(a) versus frequency (horizontal axis) and radius (vertical axis) using the permittivity values of Si in [38]. A nanosphere made of single crystalline Si can be fabricated using femtosecond LASER ablation method. One can observe the magnetic dipole-like Mie resonance as a polarizability peak in Fig. 3 ranging from shorter wavelengths for smaller particle radii to longer wavelength for larger particles. Based on formulas 4.56 and 4.57 at page 101 in [36], [37] magnetic polarizability $\alpha^{mm}$ larger than $10^{-20}$ m$^3$ are achievable with Si nanospheres with radius greater than 60 nm, where a few selected cases with $r_s = $ 60, 80, 100 nm are plotted on the right panel to better illustrate and quantify the values and line widths versus wavelength. However since nanospheres have a non negligible physical size in terms of wavelength, the force evaluation based on point-like dipole assumption should be further investigated for realistic cases with high refractive index nanospheres. The discussion here is based on having nanospheres with magnetic polarizability at optical frequency, and we consider the Si nanosphere as an example only to show the order of magnitude of the achievable polarizabilities. In the following we assume the magnetic polarizability to be $\alpha^{mm} = (-7.84 + i14.92) \times 10^{-21}$ m$^3$ at $\lambda_0 = 632$ nm.

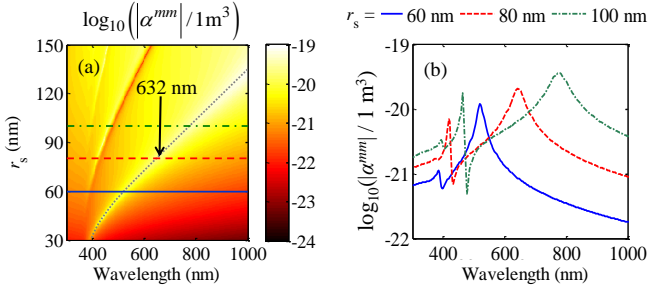

Fig. 3. (a) The logarithmic colormap of Mie polarizability $\log_{10}\left(\left|\alpha_{Si}^{mm}\right|/1m^3\right)$ of a Si nanosphere versus free-space wavelength(horizontal axis) and radius (vertical axis). The feature denoted by the gray dotted line is the 1st magnetic Mie resonance. (b) Magnetic polarizability versus wavelength for radii $r_s = 60, 80, 100$ nm.

Using such magnetic polarizability in a system as in Fig. 1, we report in Fig. 4(a) the force exerted on one of the nanoparticles according to (7) and (8). We recall that we have used the quasistatic approximation of the Green's function $G_{zz}^{Hm}$ in (6) that leads to the simple expressions (7) and (8). In Fig. 4(b) we show the magnetic force values also when all the dynamic terms of the Green's function $G_{zz}^{Hm}$ are kept in the calculation of the magnetic dipole moment and the GF derivative in (5). Results indicate that a force in the order of 0.6 pN is present in the system for $d = 210$ nm, and that the quasistatic approximation of the GF provides underestimated force values. Neglecting the coupling term in the evaluation of the magnetic dipole strength in (5), on the other hand, leads to the overestimation of the force at short distance $d$ as discussed with regard to Fig. 2. These results show that the coupling term $(1-\alpha^{mm}G_{zz}^{Hm})$ in determining $m$ must be accounted for distances smaller than 400 nm for the considered value of magnetic polarizability $\alpha^{mm}$. Interestingly, using the more accurate evaluations that include the GF dynamic terms show that the force turns from attractive to repulsive as $d$ exceeds 400nm. Also the observation that the force (5) vanishes at a certain distance close to half a wavelength [Fig. 4(b)] underlines the importance of the dynamic terms in the GF. Finally we would like to point out that high-density nanospheres with a "magnetic" Mie resonance can be polarized also electrically, i.e., they may be induced an electric polarization by the illuminating field. However illuminating vector beams with high magnetic to electric field contrast, such as vortex beams with azimuthal electric field polarization [35], that possess a vanishing electric field and longitudinal magnetic field along the propagation axis, open the scenario where at the electric force can be significantly reduced compared to the magnetic force, leaving the magnetic force the only detectable force contribution.

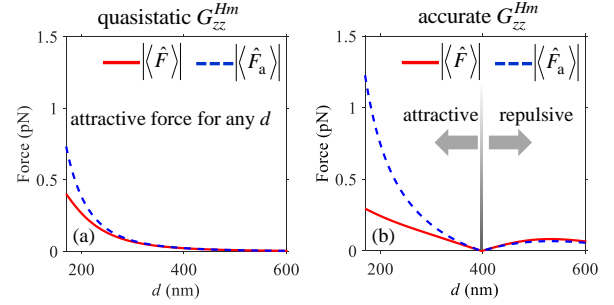

Fig. 4. The force exerted on a magnetically polarizable nanosphere, versus inter-particle distance $d$ in a two particle system as in Fig. 1 at 632 nm wavelength with a magnetic polarizability of $\alpha^{mm} = (-7.84+i14.92)\times10^{-21}$ m$^3$. Using the quasistatic approximation of $G_{zz}^{Hm}$ the force is attractive as in (a). When the dynamic terms of $G_{zz}^{Hm}$ are included as in (b) at short distance the magnetic force is stronger whereas at larger distances the force is repulsive.

### III. FORCE BETWEEN TWO MAGNETICALLY POLARIZABLE PLASMONIC NANOCLUSTERS

We investigate the possibility of force detection using artificial magnetic dipoles made of circular clusters of plasmonic nanoparticles as in Fig. 5(a) that, for simplicity, is considered non deformable. Such plasmonic-based magnetic clusters can be sculpted on a microscope nanotip using focused-ion beam. For this investigation a cluster of plasmonic nanospheres shall suffice to demonstrate the feasibility of magnetic force detection. Similarly to the previous case, we assume that the two circular clusters are immersed in a time harmonic field at a frequency close to the loop-like resonance of a circular cluster. We evaluate the time-average force associated to the interaction of near magnetic fields and the cluster resonance that supports circulating plasmons. However, since each resonant loop constitutes an effective magnetic dipole, this force interaction can also be equivalently interpreted as the one caused by equivalent magnetic dipoles immersed in a magnetic field as in the previous section.

Let us consider a system made of $N$ polarizable plasmonic nanoparticles at $\mathbf{r}_n$, with $n=1,\ldots,N$. The equivalent electrical dipole moment of each particle is determined by $\mathbf{p}_n = \underline{\alpha}_n \cdot \mathbf{E}^{loc}(\mathbf{r}_n)$ where $\underline{\alpha}_n^{ee}$ is the electric polarizability tensor and

$$\mathbf{E}_n^{loc} = \mathbf{E}^{ext}(\mathbf{r}_n) + \sum_{m\in\{1,\ldots,N\}\setminus\{n\}} \underline{\mathbf{G}}^{Ep}(\mathbf{r}_n - \mathbf{r}_m)\cdot \mathbf{p}_m \qquad (9)$$

is the local electric field at the $n$th dipole position generated by the external electric field excitation and the electric field due to all the other polarized particles in the system [made of either one or two clusters as in Fig. 5 (b) ]. Here $\underline{\mathbf{G}}^{Ep}$ is the dyadic GF used for evaluating the electric field at $\mathbf{r}_n$ due to an electric dipole at $\mathbf{r}_m$ with a given moment $\mathbf{p}_m$. Then by multiplying (9) by $\underline{\boldsymbol{\alpha}}_n^{ee}$ we obtain the system of N equations

$$\mathbf{p}_n - \underline{\boldsymbol{\alpha}}_n^{ee} \sum_{m \in \{1,..,N\}\setminus\{n\}} \underline{\mathbf{G}}^{Ep}(\mathbf{r}_n - \mathbf{r}_m) \cdot \mathbf{p}_m = \underline{\boldsymbol{\alpha}}_n^{ee} \mathbf{E}^{ext}(\mathbf{r}_n) \quad (10)$$

where $n=1,\ldots N$, that can be solved for the $N$ dipole moments $\mathbf{p}_n$. As we are using nanospheres in the following, the electric polarizability reduces to the scalar polarizability $\alpha^{ee}$, here evaluated by Mie theory [36]. When excited at the so-called "magnetic resonance" frequency [6], [13], the cluster assumes a mainly circulating disposition of dipole moments, as in Fig. 5(a), that forms an equivalent magnetic dipole.

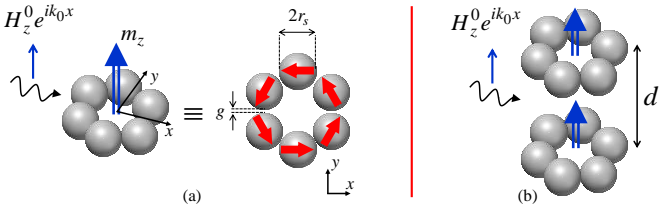

Fig. 5. (a) A nanocluster exhibiting an effective magnetic dipole moment normal to the cluster plane generated by a circulating electric dipole disposition. (b) The two-cluster system similar to the one in Fig. 1, for which the force is evaluated.

In a frequency range around the magnetic resonance the field along the z-axis is strongly enhanced as reported in [13], and lead to force manipulation. Next we show that a circular cluster as in Fig. 5(a) possesses magnetic polarizability levels shown to yield desired force levels as evaluated in Fig. 2. For this purpose we consider a circular nanocluster made of 6 identical silver nanospheres in vacuum excited by a plane wave propagating in the $x$ direction, with magnetic field $H_z^{ext} = H_z^0 e^{ik_0 x}$ polarized along $z$, as depicted in Fig. 6(a). The equivalent magnetic dipole moment of the cluster is given by

$$\mathbf{m} = \frac{-i\omega}{2} \sum_{\text{Cluster}} \mathbf{r}_n \times \mathbf{p}_n \quad (11)$$

where the electric dipole moments $\mathbf{p}_n$ are determined by solving (9). The $zz$ entry of the overall equivalent *magnetic* polarizability tensor of this nanocluster is defined via

$$\alpha_{zz}^{mm} = \frac{m_z}{H_z^0} \quad (12)$$

that is reported in Fig. 6 for the several nanosphere radii and gap distances varied around the central design with a radius of 50 nm and a gap of 7.5 nm. Nanospheres are assumed to be made of silver with measured permittivity function taken from [38]. In the considered ranges, the polarizability level and linewidth is mainly controlled by the nanosphere radius, rather than the gap distance as observed in Fig. 6. Importantly, these plots show that magnetic polarizability values in the order of $10^{-20}$ m$^3$ are realizable, close to the required values reported in Sec. II.

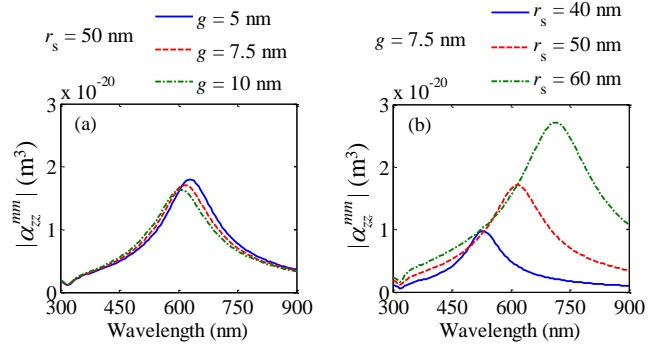

Fig. 6. The magnitude of the $zz$ entry of the magnetic polarizability tensor of circular cluster of silver nanospheres (left) for various gap distances $g$ and (right) for various nanosphere radii $r_s$. The obtained values are comparable to those in Fig. 3.

When two such nanoclusters are placed close to each other and illuminated with a plane wave as illustrated in Fig. 5(b), a strong magnetic field is generated at the magnetic resonance which, in turn, exerts a force on each cluster. The instantaneous magnetic force exerted on a cluster is evaluated with the magnetic component of the Lorentz force

$$\hat{\mathbf{F}}(t) = \mu_0 \int \left[ \hat{\mathbf{J}}(\mathbf{r}) \times \hat{\mathbf{H}}(\mathbf{r}) \right] d\mathbf{r}^3 \quad (13)$$

where the integral domain coincides with that of all the nanospheres of a cluster. Here $\hat{\mathbf{J}}$ is the volume current density associated to the electron movement in the plasmonic nanospheres, and $\hat{\mathbf{H}}$ is the time domain magnetic field. The volume of each polarizable nanoparticle is subwavelength and close to its plasmonic resonance the current inside each particle is basically uniform because of the spherical shape [36] leading to $d\hat{\mathbf{p}}_n / dt = \int \hat{\mathbf{J}}(\mathbf{r}) d\mathbf{r}^3$. Under point dipole limit, the magnetic component of the Lorentz force (13) on each subwavelength-size electrically polarizable particle is solely due to the *local* magnetic field [39] which varies slowly at the equivalent dipole position (i.e. the center of the subwavelength nanoparticle volume). Then the above magnetic force integral reduces to the summation

$$\hat{\mathbf{F}}(t) = \mu_0 \sum_{n \in \text{Cluster}} \frac{d\hat{\mathbf{p}}_n}{dt} \times \hat{\mathbf{H}}^{loc}(\mathbf{r}_n). \quad (14)$$

We recall that we assume the clusters to be non deformable, and we are interested in the force exerted on one of the clusters by the other one, therefore the summation in (14) is over a single cluster (note the symmetry in the system), whereas local magnetic field $\hat{\mathbf{H}}^{loc}$ is taken as the field produced by the other cluster and the external beam. Assuming time harmonic excitation, the instantaneous dipole moments are represented as $\hat{\mathbf{p}}_n = \text{Re}\{\mathbf{p}_n e^{-i\omega t}\}$. Therefore, analogously to what we have shown at the beginning of this paper, the instantaneous force has the form

$$\hat{\mathbf{F}}(t) = \frac{\omega \mu_0}{2} \sum_{n \in \text{Cluster}} \text{Im}\{\mathbf{p}_n \times \mathbf{H}_n^{\text{loc}*}\} + \frac{\omega \mu_0}{2} \sum_{n \in \text{Cluster}} \text{Im}\{\mathbf{p}_n \times \mathbf{H}_n^{\text{loc}} e^{-i2\omega t}\} \quad (15)$$

where the first term represents the time-average magnetic force $\langle \hat{\mathbf{F}} \rangle$ that will be reported in Fig. 7. Note that when the ring of particles hosts circulating dipolar currents, the local magnetic field and the dipole moments orthogonal to the $z$ axis lead to net force along the $z$ direction on both clusters. When currents are slightly distorted from a perfect loop-like distribution also other force components may occur. Both the local magnetic field and the induced dipole moments are proportional to the incident electric field amplitude, thus the force is linearly proportional to the incident beam's intensity. This force will be denoted to as "magnetic force" since it is generated by the interaction with the magnetic field through the $\hat{\mathbf{J}} \times \hat{\mathbf{B}}$ term of the Lorentz force (in contrast to the electric force $\hat{\mathbf{p}} \cdot \nabla \hat{\mathbf{E}}$ exerted to an electric dipole in an electric field $\hat{\mathbf{E}}$). The magnetic force between the two clusters here is analogous to the force between of two magnetically polarizable nano particles in Sec II.

Consider a plane wave (i.e, with average power density of $1.33 \times 10^5$ W/cm$^2$), illuminating the two circular nanoclusters, each made of 6 identical nanospheres with a radius of 50 nm and a gap of 7.5 nm, with the magnetic field amplitude of $2.65 \times 10^3$ A/m as in Sec II. In the following the force is evaluated by (15), even though in Fig. 6 we reported the strength of the magnetic polarizability for establishing the analogy with the magnetic dipoles. The $z$-component of the magnetic force exerted on a cluster by the other one is reported in Fig. 7, versus wavelength and inter-cluster distance $d$. Note that the $z$-directed force is purely due to the magnetic near field of the two clusters because the incident magnetic field along the $z$ axis does not exert a force along the $z$ direction as implied by cross-product in (15). We consider the $H$ field produced by the other cluster only in Fig. 7 and we neglect the one produced by the cluster itself. In Fig. 7(a), there is a common peak around 632 nm wavelength for different inter-cluster distances, while other peaks are observed at smaller wavelength ranges, out of the magnetic resonance band. The peaks around 632 nm correspond to a circular disposition of electric dipole moments, while others are "magnetic-like" resonances excited due to the inter-cluster interaction. In Fig. 7(b), the force profile versus the inter-cluster distance is plotted for three wavelengths around 632 nm. Note that calculations yield force levels in the order of 0.3 pN. Moreover the dependence of force on the inter-cluster distance shows remarkable resemblance to the case reported in Fig. 4(b). For the sake of comparison, in Fig. 8 the average electric force

$$\langle \hat{\mathbf{F}}^e \rangle = \frac{1}{2} \sum_{n \in \text{Cluster}} \text{Re}\{\mathbf{p}_n \cdot [\nabla \mathbf{E}_n^{\text{loc}}]^*\} \quad (16)$$

is reported for the same cases as in Fig. 7. Remarkably, the electric force is in general in the opposite direction of the magnetic force and it is up to 7.5 times weaker than the magnetic force around the magnetic resonance frequency in the reported distance range when $d > 150$ nm. This implies that the overall force direction is dominated by the magnetic force, therefore the force measurement can be efficiently used for detecting the magnetic resonance. The electric force can be further suppressed by exciting the cluster with structured beams such as azimuthally polarized vector beams [35] with longitudinal magnetic field and vanishing electric field along the propagation axis. We also remind that the inner force acting on a cluster is not necessarily zero. This does not violate the conservation of momentum, because the total momentum is given by the sum of electromagnetic and mechanical momenta [40]. For example when the power is scattered in a directional manner, the particle gains momentum in the opposite direction. Here a cluster is thus also exerting a force also onto itself when in (15) the magnetic field contributions by the cluster itself are considered, but the strength is much weaker than what reported in Fig. 7 around the magnetic resonance at 632 nm. The results obtained by dipolar scattering assumption is known to deviate from the accurate full-wave representation when the gap between nanospheres is smaller than a radius. However the force levels reported here occur even when the polarizable particles have gaps larger than a diameter, within the reliability range of the employed model.

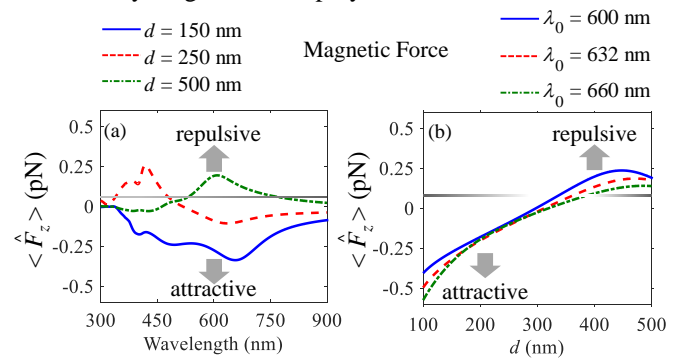

Fig. 7. The $z$-component of the magnetic force exerted on the upper cluster by the bottom one (left) for various inter-cluster distances versus wavelength, and (right) for various wavelengths versus inter-cluster distance. In all cases the nanospheres are identical with a radius equal to $r_s$ = 50 nm and a gap equal to $g$ = 7.5 nm.

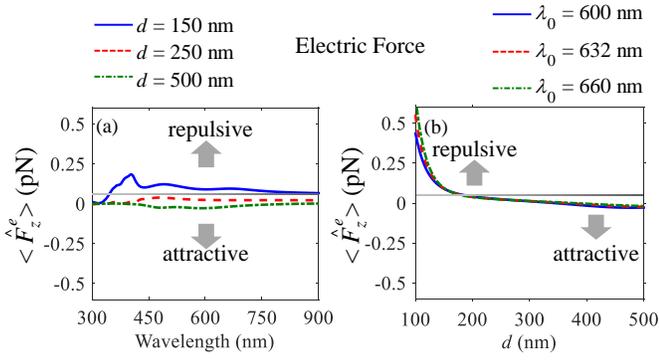

Fig. 8. The *z*-component of the average electric force for the same case in Fig. 7.

## IV. CONCLUSION

We have shown the photo-induced magnetic force interaction between two nanostructures with magnetic polarizability at optical frequency is in the range of a piconewton. Therefore in nanostructures where optical magnetism is properly enhanced, magnetic forces can be detected in principle through a scanning force microscope. Magnetic dipolar transitions, like singlet-triplet transitions, are dark in current electro-optical technologies, and thus remain relatively unexplored in optical spectroscopy. It is foreseen that the conclusions drawn here pave the way for utilizing force microscopy for the investigation of artificial magnetism and weak magnetic-dipole transitions in matter manifesting themselves at optical frequencies. Magnetic dipole transitions in matter can be properly enhanced using magnetic nanoprobes as those shown here and by using structured light as in [35] where the incident beam is dominated by the magnetic field component. These transitions may be detected by a scanning probe microscope with nanoscale precision dictated by the magnetic nanoprobe near-field hot spot.

## V. ACKNOWLEDGMENT

The authors are thankful to the W. M. Keck Foundation for its generous support.

## APPENDIX

In Sec. II, the magnetic force between two magnetically polarizable nanoprobes positioned along the *z* axis under *z*-polarized magnetic field was investigated. It is also interesting to discuss on how the force evolves if these two nanosphere are misaligned with a relative displacement $\Delta$ along the *y* axis. Since the symmetry with respect to the *z* axis breaks down in this case, the induced magnetic dipole and the magnetic field interaction between them cannot be reduced to simple scalar equations. Therefore a system solution of the magnetically polarizable particles under magnetic field excitation analogous to the electric case summarized in equations (9) and (10) is employed. Namely one has to solve for the two magnetic dipole moments through the following two vector equations

$$\mathbf{m}_i - \alpha^{mm} \sum_{j \in \{1,..2\} \setminus \{i\}} \underline{\mathbf{G}}^{Hm}(\mathbf{r}_i - \mathbf{r}_j) \cdot \mathbf{m}_j = \alpha^{mm} \mathbf{H}^{ext}(\mathbf{r}_i) \quad (17)$$

where $i = 1, 2$.

The average magnetic force calculation is carried out using the formula

$$\langle \hat{\mathbf{F}} \rangle = \frac{\mu_0}{2} \operatorname{Re} \left\{ \mathbf{m} \cdot \left[ \nabla \mathbf{H}^{loc} \right]^* \right\}. \quad (18)$$

In Fig. 9, we report the *z* and *y* components of the average magnetic force $\langle \hat{\mathbf{F}} \rangle$ on the upper nanoprobe versus the vertical distance *d*, for various center-to-center distances $\Delta$ along the *y* direction (other physical parameter are the same as those considered in Fig. 4). It is observed that the *z*-directed force reaches attractive maximum when the two nanospheres are aligned along the *z* axis. When the nanoprobes are displaced along the *y* direction, the magnetic force $\langle \hat{F}_z \rangle$ significantly reduces when $\Delta$ reaches only 150 nm. (note that a Si nanosphere with the same magnetic polarizability considered in this example would have a diameter of 160 nm). The displacement along the *y* direction also causes a *y*-directed force $\langle \hat{F}_y \rangle$, which is attractive at small vertical distances, but it is in general significantly smaller than the *z*-directed force. In Fig. 10, the *z* and *y* components of magnetic force is plotted versus the horizontal center-to-center distance $\Delta$ for various *d*. The *z* component of the force is attractive for small $\Delta$ and turns repulsive for larger $\Delta$ when $d = 160, 250, 350$ nm. Moreover Fig. 10 shows that the degree of spatial resolution based on both the force level and polarity that can be achieved varying the displacement $\Delta$, especially for shorter distances *d*. However for $d = 450$ nm the magnitude and polarity of the force do not provide the good resolution. The *y* component of the force is mainly attractive and reaches comparable levels to the *z* component for large $\Delta$.

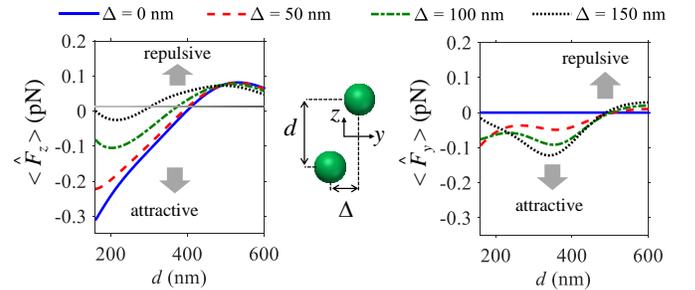

Fig. 9. The *z* and *y* component of the magnetic force between two magnetically polarizable nanosphere versus the vertical center-to-center distance for various horizontal distances $\Delta$ along the *y* direction. All other parameters are as in Fig. 4.

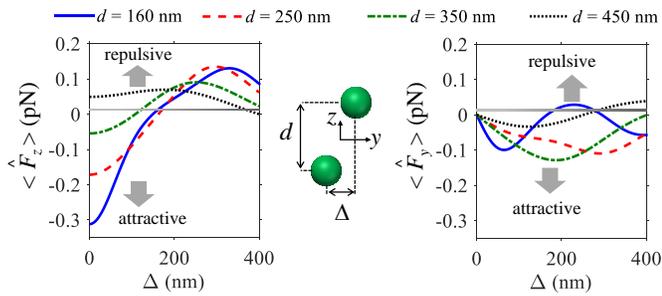

Fig. 10. The z and y component of the magnetic force between two magnetically polarizable nanosphere versus the horizontal center-to-center distance $\Delta$, for various $d$. All other parameters are as in Fig. 4.